\documentclass[a4paper,12pt]{article}

\def\displayandname#1{\rlap{$\displaystyle\csname #1\endcsname$}%
                      \qquad \texttt{\char92 #1}}

\usepackage{epsfig}

\begin{document}

\title{Sakata model of hadrons revisited}

\author{Eugene V. Stefanovich \and \small \emph{2255 Showers Drive, Unit
153, Mountain View, CA 94040, USA} \and \small
$eugene\_stefanovich@usa.net$}

\maketitle

\begin{abstract}
46 years ago the quark model replaced the Sakata model
as the standard explanation of the hadron structure. The major alleged defect of the Sakata model was its prediction of just too many types of particles, which have not been seen in experiments. However, this allegation was made without detailed consideration of the forces acting between sakatons. In this article we suggest a set of pairwise sakaton-sakaton and sakaton-antisakaton potentials that describe stability and masses of strongly interacting elementary particles in a good agreement with observations.
\end{abstract}

\section{ The Sakata model} \label{sc:sakata}

 Today it is
universally accepted that hadrons are made of quarks ($q = u,d,s,c,\ldots$). The quark model forms the basis of quantum chromodynamics (QCD), which aspires to explain the nature of strong interactions. Almost all compound particles predicted by the quark model have been found in experiments. Moreover, all observed particles have natural quark assignments: Mesons are quark-antiquark bound state ($q\overline{q}$), and
baryons are bound states of three quarks ($qqq$).\footnote{Some suggested tetraquark and pentaquark assignments are not universally accepted. All experimental data about hadron properties used in this paper were taken from  \cite{PDBook}. Experimental values are emphasized by the bold font in this paper.} In spite of their
well-known achievements, the quark model and QCD have some questionable
features. These theories make assumptions (fractional charges of quarks, color, gluons, confinement potentials, etc.), which cannot be directly observed and thus destined to remain suspect.  Then it seems justified to explore other approaches to the explanation of hadrons masses and stability. One interesting proposal is the Sakata model \cite{Sakata}, which was rather popular before the ``quark era''. The Sakata model assumes that
proton ($p$), neutron ($n$), $\Lambda^0$ and $\Lambda_c^+$  are the true elementary particles\footnote{In this paper we will not discuss bottom and top particles, because full experimental picture is still lacking in those sectors.} also called \emph{sakatons} ($\sigma$) \cite{sakaton}.  To emphasize their similarity with
quarks, we will denote the four fundamental sakatons by capital letters
$U,D,S,C$.\footnote{See  Table \ref{table:4.1d}. We use symbol $N$ do denote
collectively $U$ and $D$ sakatons. For example, $NN\overline{S}$ means either $UU\overline{S}$ or $DD\overline{S}$.}  Each sakaton has its corresponding antisakaton
($\overline{U},\overline{D},\overline{S},\overline{C}$)
with the same mass and spin and opposite values of the electric
charge, baryon charge, strangeness, and charm.

\begin{table}[h]
\caption{Stable baryons with their quark and sakaton structures.}
\begin{tabular*}{\textwidth}{@{\extracolsep{\fill}}lccccccc}
 \hline
baryon & Quark     &   sakaton   & Exp. mass\cr
       & structure &  structure &     MeV/$c^2$ \cr
 \hline
$n$        & $udd$ &  $D$ (down)                       & \textbf{938}     \cr
$p$        & $uud$ &  $U$ (up)                      & \textbf{940}     \cr
$\Lambda^0$ & $sud$ &  $S$ (strange)                      & \textbf{1116}     \cr
$\Lambda_c^+$ & $cud$ &  $C$ (charmed)                      & \textbf{2285}     \cr
\hline
$\Sigma^-$ &$sdd$ &   $SD\overline{U}$          & \textbf{1197}     \cr
$\Sigma^0$ &$sud$ &  $SN\overline{N}$   & \textbf{1193}   \cr
$\Sigma^+$ &$suu$ &  $SU\overline{D}$         & \textbf{1189}   \cr
$\Xi^-$ &   $ssd$ &   $SS\overline{U}$        & \textbf{1322}  \cr
$\Xi^0$ &   $ssu$ &   $SS\overline{D}$            & \textbf{1315} \cr
$\Xi_c^0$  & $csd$ &  $CS\overline{U}$           & \textbf{2471} \cr
$\Xi_c^+$  & $csu$ &  $CS\overline{D}$           & \textbf{2468}  \cr
$\Xi_{cc}^{+}$ & $ccd$ & $CC\overline{U}$        & \textbf{3519}\cr
$\Xi_{cc}^{++}$ & $ccu$ &  $CC\overline{D}$       & not seen \cr
\hline
$\Omega^-$ & $sss$ &  $SSS\overline{U}\overline{D}$  & \textbf{1672}\cr
$\Omega_c^0$ & $css$ &  $CSS\overline{U}\overline{D}$  & \textbf{2698}\cr
$\Omega_{cc}^{+}$  & $ccs$ &  $CCS\overline{U}\overline{D}$   & not seen \cr
$\Omega_{ccc}^{++}$  & $ccc$ &  $CCC\overline{U}\overline{D}$   & not seen \cr
\hline
\end{tabular*}
\label{table:4.1d}
\end{table}

The Sakata model assumes that sakatons interact with each other via short-range (few femtometers) potentials. All non-elementary hadrons are bound states of two or more sakatons. Various possible combinations are summarized in Table \ref{table:10}. Nuclei are composed of $U$ and $D$ sakatons (protons and neutrons).   Mesons  are sakaton-antisakaton ($\sigma\overline{\sigma}$) bound states.  Compound baryons are
sakaton-sakaton-antisakaton ($\sigma\sigma\overline{\sigma}$) or \emph{pentasakaton} ($\sigma\sigma\sigma\overline{\sigma}\overline{\sigma}$) bound
states.  Here we are interested only in baryons, which are stable with respect to strong decays. All of them are listed in Table \ref{table:4.1d}. Their decays are caused by flavor-changing weak interactions and their masses are lower than the sums of masses of constituents.    One example of an unstable baryon state omitted in Table \ref{table:4.1d} is the $\Delta^{++} (=UU\overline{D})$ particle whose mass is \textbf{1232} MeV/$c^2$. This is higher than the sum of masses of dissociation products $p(\textbf{940}) + \pi^+ (\textbf{140})(= U + U\overline{D})$. Therefore $\Delta^{++}$  is a metastable resonant state in our model. The calculation method adopted in this work (see section \ref{sc:model}) can deal only with true bound states, therefore we will not discuss the $\Delta^{++}$  and other resonances.

 The Sakata model avoids some problems characteristic for
the quark model. The fundamental constituents of the Sakata model --
the sakatons -- are readily observable as normal baryons with integer charges, so
there is no need for additional assumptions about "confinement".
There is also no need to introduce "hidden" degrees of freedom, such
as color and gluons. The short-range character of sakaton potentials means that strong interactions satisfy the important property of cluster-separability \cite{book}, similar to electromagnetic and gravitational forces.

\begin{table}[h]
\caption{Bound states of sakatons}
\begin{tabular*}{\textwidth}{@{\extracolsep{\fill}}llll}
 \hline \hline
sakaton  content & particle type &  examples  & antiparticle  \cr
 \hline
 $\sigma$ & baryon  & $p,n, \Lambda^0, \Lambda_c^+$ & $\overline{\sigma}$ \cr
  $N N' \ldots N''$ & nucleus & deuteron($=UD$) & $\overline{N} \overline{N}'
\ldots \overline{N}''$ \cr
  $\sigma \overline{\sigma}$ & meson & $K^+(=U\overline{S})$ & $\sigma \overline{\sigma}$ \cr
  $\sigma \sigma \overline{\sigma}$ & baryon  & $\Sigma^{-}(=SD\overline{U})$ &
 $\sigma \overline{\sigma} \overline{\sigma}$ \cr
 $\sigma \sigma \sigma \overline{\sigma}$ & tetrasakaton  & unstable? &
$\sigma \overline{\sigma} \overline{\sigma} \overline{\sigma}$ \cr
 $\sigma \sigma \overline{\sigma} \overline{\sigma}$ &  tetrasakaton & unstable? &
$\sigma \sigma \overline{\sigma} \overline{\sigma}$ \cr
 $\sigma \sigma \sigma \overline{\sigma} \overline{\sigma}$ & baryon &
$\Omega^-(=SSS\overline{U}\overline{D})$ &
 $\sigma \sigma \overline{\sigma} \overline{\sigma}
\overline{\sigma}$ \cr
 \hline
\end{tabular*}
\label{table:10}
\end{table}

The biggest problem of the Sakata model is that it seemingly predicts more types
of particles than actually observed.
 Certain $\sigma\sigma\overline{\sigma}$ combinations, which
look acceptable from the point of view of the Sakata model, have not
been seen in experiments. This refers, for example to
$NN\overline{S}$ and $UD\overline{S}$ baryons with strangeness +1.\footnote{Reports about discovery of the exotic baryon $\Theta^+(=UD\overline{S})$ are not credible \cite{PDBook}.} Furthermore, the simplest sakaton assignment of the $\Omega^-$ baryon (baryon
number = 1, charge = -1, strangeness = -3) is in the form of a
pentasakaton  $\Omega^-(=SSS\overline{U}\overline{D})$. Then, from the principle of isotopic invariance, it seems that analogs of the $\Omega^-$ particle should also exist, such as $\Omega^{--}(= SSS\overline{U}\overline{U})$ and $\Omega^{0}(= SSS\overline{D}\overline{D})$. Why haven't they been seen in experiments?

In order to answer these and other questions, it is important  to have a realistic
model of interactions between sakatons. The goal of this paper is to suggest an approximate set of pairwise sakaton-sakaton and sakaton-antisakaton potentials  and to calculate masses of their bound states - mesons and baryons.

\section{Computational model and results}
\label{sc:model}

Matumoto and co-authors established \cite{Matumoto-56,
Matumoto-60, Matumoto-61, Matumoto-61a, Sawada} that masses of hadrons can be roughly calculated from the assumption of strong attraction in sakaton-antisakatons pairs (i.e., one $\sigma - \overline{\sigma}$ bond contributes about 1275-1740 MeV to the binding energy) and equally strong sakaton-sakaton and antisakaton-antisakaton repulsions. Binding energies of mesons are very high (above 1GeV), because only the $\sigma - \overline{\sigma}$ attraction contributes there. Much lower binding energies are expected in 3-sakaton $\sigma \sigma \overline{\sigma}$  and in pentasakaton $\sigma \sigma \sigma \overline{\sigma} \overline{\sigma}$ baryons. In the former case two attractive interactions $\sigma - \overline{\sigma}$ are balanced by one repulsion $\sigma -\sigma$. In the latter case there are 6 attractions vs. 4 repulsions. Tetrasakatons $\sigma\sigma\sigma \overline{\sigma}$ are not likely to be stable because the number of repulsive and attractive pairs is equal in this case.  Some instructive studies of multiparticle systems with pairwise interactions can be found in \cite{Martin, Armour}. They suggest that stability of multi-sakaton states may depend on a delicate balance of masses of the constituents and shapes of their interaction  potentials.

The approximate non-relativistic Hamiltonian describing an
$\mathcal{N}$-sakaton system can be written as

\begin{eqnarray}
H= \sum_{i=1}^{\mathcal{N}} m_ic^2  + \sum_{i=1}^{\mathcal{N}}\frac{p_i^2}{2m_i}  +
\sum_{i<j}^{\mathcal{N}} V_{ij}(r_{ij}) \label{eq:hamiltonian}
\end{eqnarray}

\noindent where $m_i, \mathbf{p}_i, r_{ij} = |\mathbf{r}_i -
\mathbf{r}_j|$ are masses and momenta of the sakatons and their relative distances, respectively. Interactions between sakatons were modeled as superpositions of two Yukawa potentials

\begin{eqnarray}
V_{ij}(r)= A_{ij}z_iz_j\frac{ e^{-\alpha_{ij}r}}{ r} + B_{ij}\frac{
e^{- \beta_{ij}r}}{ r} \label{eq:potential}
\end{eqnarray}

\noindent where  $z_i=+1$ for sakatons and $z_i = -1$ for antisakatons.

All calculations were performed using the stochastic variational method of Varga and Suzuki \cite{Varga, Varga2, Varga-book}. The FBS computer program was obtained from the CPC Program Library (Queen's University of Belfast, N. Ireland) and slightly modified to fit our needs. This program solves the non-relativistic stationary Schr\"odinger equation and yields accurate energies and wave functions of the ground and few excited states for systems of several (typically, 2-6) quantum particles interacting via pairwise potentials.
Only states with the lowest total spin ($s=0$ for mesons and $s=1/2$ for baryons) and zero orbital momentum were considered here. In all calculations masses of sakatons were fixed as $m(N) = 940$ MeV/$c^2$,\footnote{The equality of masses of the $U$ and $D$ sakatons and the assumption that their interactions with other sakatons are the same (see Table \ref{table:11b}) imply that all calculated masses are invariant with respect to  replacements in which all $U$ sakatons are changed to $D$ and all $D$ sakatons are simultaneously changed to $U$.} $m(S) = 1116$ MeV/$c^2$, $m(C) = 2285$ MeV/$c^2$. Internally in the code these masses were expressed in units of the proton mass 940 MeV/$c^2$. Distances were measured in femtometers and energies in MeV. In this system of units $\hbar^2/m = 41.47$. The basis set selection procedure used iteration numbers $M_0 = 10, K_0 = 50$. Other computational parameters depended on the number of sakatons in the system as shown in Table \ref{table:11x}. They were adjusted for the optimal balance between accuracy, convergence, and speed. The exact meaning of these parameters was explained in \cite{Varga}.

\begin{table}[h]
\caption{Computational parameters for the FBS code. $b_{min}/b_{max}$ are minimum/maximum values of nonlinear parameters in Gaussian basis functions.}
\begin{tabular*}{\textwidth}{@{\extracolsep{\fill}}c|ccc}
 \hline \hline
 Number of   &  Basis set  & $b_{min}$   & $b_{max}$      \cr
 sakatons   & size, $K$   &  (fm)           &   (fm)   \cr
 \hline \hline
 2           & 50  & $10^{-6}$ & 10     \cr
 3           & 250 & $10^{-6}$ & 10     \cr
 4           & 300 & $10^{-6}$ & 10      \cr
 5           & 500 & $10^{-6}$ & 100    \cr
 \hline \hline
\end{tabular*}
\label{table:11x}
\end{table}

\begin{table}[h]
\caption{Optimized parameters of the potentials (\ref{eq:potential}). $A$ and $B$ are measured in MeV$\cdot$fm; $\alpha$ and $\beta$ are in fm$^{-1}$.}
\begin{tabular*}{\textwidth}{@{\extracolsep{\fill}}r|cccc}
 \hline \hline
 Interaction            &  $A$  & $\alpha$   & $B$   &  $\beta$     \cr
 \hline \hline
 $N-N$     & 617.8 & 0.091 & 92.14 & 0.359    \cr
  $U-D$              & 570.2 & 0.091 & 25.7  & 0.094    \cr
 $N-S$      & 530.0 & 0.108 & 14.0  & 0.49     \cr
  $ S-S$             & 446.7 & 0.118 & 42.1  & 0.444    \cr
    $N-C$     & 397.5 & 0.102 & 14.0  & 0.49     \cr
  $S-C$              & 340.8 & 0.12  & 46.1  & 0.444     \cr
  $C-C$              & 317.0 & 0.118 & 24.1  & 0.484    \cr
 \hline \hline
\end{tabular*}
\label{table:11b}
\end{table}

\begin{figure}
\epsfig {file=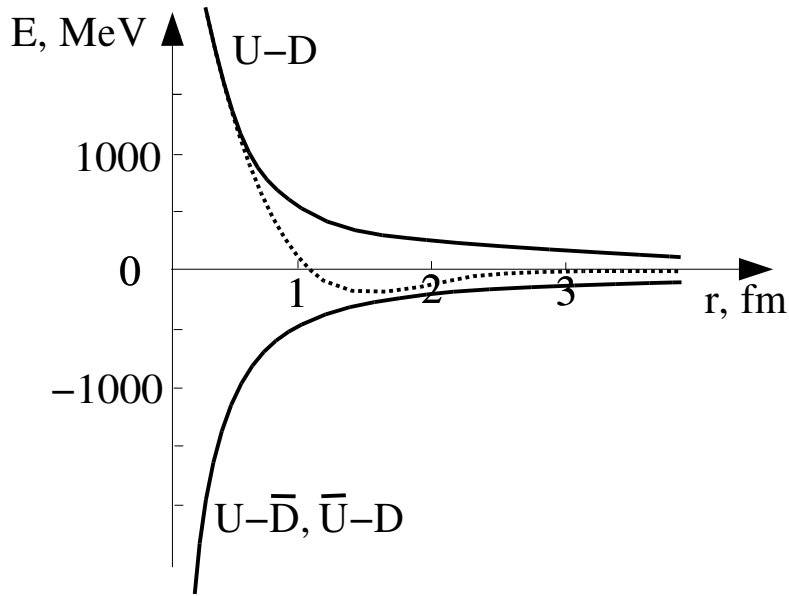} \caption{Schematic shape of the
$U-D$, $U-\overline{D}$ and $\overline{U}-D$ interaction potentials (full lines).
The broken line shows a proposed modification of the $U-D$ potential that can explain the bonding in nuclei. } \label{fig:3}
\end{figure}

Our major goal is to optimize parameters $A, \alpha, B, \beta$ of the potentials (\ref{eq:potential}). The optimization was performed in two steps. In the first step we fitted parameters relevant to interactions of $U,D,S$ sakatons. The training set contained 24 species shown in Table \ref{table:6c}. They included 4 ground states of mesons, 4 stable baryons, and 16 states, which are supposed to be unstable. The goal was to reproduce experimental masses of the 8 stable species as close as possible and, at the same time, do not allow the binding energy of the 16 unstable species to become positive.  In the second step we froze the $U-D-S$ parameters obtained above and varied interactions $C-N$, $C-S$, and $C-C$ using the training set in Table \ref{table:6d}. This set included $C$-containing particles: 3 mesons, 3 stable charmed baryons, and 17 unstable species. The final optimized values of parameters $A, \alpha, B, \beta$ are given in Table \ref{table:11b}. Plots of the optimized $U-D$ and $U-\overline{D}$ (same as $\overline{U}-D$) potentials are shown in Fig. \ref{fig:3}. Potentials for other pairs of sakatons have qualitatively similar shapes. These interactions demonstrate rather strong attraction of $\sigma - \overline{\sigma}$ pairs and repulsion of $\sigma - \sigma$ and $\overline{\sigma} - \overline{\sigma}$ pairs in a qualitative agreement with Matumoto's guesses.

\begin{table}[h]
\caption{Training set for no-charm particles.}
\begin{tabular*}{\textwidth}{@{\extracolsep{\fill}}lllcc}
 \hline
 Particle & Sakaton & Mass (MeV/$c^2$) & B.E. (MeV) & Products   \cr
         &   structure & calc./\textbf{exp.} & calc./\textbf{exp.} & calc./\textbf{exp.} \cr
 \hline
 $\pi^0$ & $D\overline{D}$ & 238/\textbf{135}  & 1642/\textbf{1745} & $n+\overline{n}$/$\mathbf{n}+\overline{\mathbf{n}}$ \cr
 $\pi^+$ & $U\overline{D}$ & 142/\textbf{141}  & 1738/\textbf{1739} & $p+\overline{n}$/$\mathbf{p}+\overline{\mathbf{n}}$ \cr
 $K^-$   & $S\overline{U}$ & 364/\textbf{494}  & 1692/\textbf{1562} & $\overline{p}+\Lambda^0$/$\overline{\mathbf{p}}+\mathbf{\Lambda}^0$ \cr
  $\eta$  & $S\overline{S}$ & 1095/\textbf{548} &1137/\textbf{1684}  & $\Lambda^0+\overline{\Lambda}^0$/$\mathbf{\Lambda}^0+\overline{\mathbf{\Lambda}}^0$ \cr
\hline
  $\Sigma^+$& $SU\overline{D}$ & 1210/\textbf{1189} & 48/\textbf{67} & $\Lambda^0+\pi^+$/$\mathbf{\Lambda}^0+\mathbf{\pi}^+$ \cr
 $\Sigma^0$& $SU\overline{U}$ & 1260/\textbf{1193} & 44/\textbf{58} & $p+K^-$/$\mathbf{\Lambda}^0+\mathbf{\pi}^0$ \cr
  $\Xi^0$ & $SS\overline{D}$ & 1314/\textbf{1315}        & 166/\textbf{295} & $\Lambda^0+K^0$/$\mathbf{\Lambda}^0+\mathbf{K}^0$ \cr
  $\Omega^-$& $SSS\overline{U}\overline{D}$ & 1670/\textbf{1672}  & 8/\textbf{136} & $\Xi^-+K^0$/ $\mathbf{\Xi}^-+\mathbf{K}^0$ \cr
\hline
unstable    & $UU\overline{U}$ & 1179        & -1/\textbf{0} & $p + \pi^0$/$\mathbf{p} + \mathbf{\pi}^0$ \cr
unstable    & $UD\overline{U}$ & 1082        & 0/\textbf{0} & $p + \pi^-$/$\mathbf{p} + \mathbf{\pi}^-$ \cr
unstable          & $UU\overline{S}$ & 1307        & -3/\textbf{0} & $p + K^+$/$\mathbf{p} + \mathbf{K}^+$ \cr
unstable          & $UD\overline{S}$ & 1307        & -3/\textbf{0} & $p + \overline{K}^0$/$\mathbf{p} + \overline{\mathbf{K}}^0$ \cr
unstable   & $SU\overline{S}$ & 1482        & -2/\textbf{0} & $\Lambda^0 + K^+$/$\mathbf{\Lambda}^0 + \mathbf{K}^+$ \cr
unstable          & $SS\overline{S}$ & 2211        & 0/\textbf{0} & $\Lambda^0 + \eta$/$\mathbf{\Lambda}^0 + \mathbf{\eta}$ \cr
unstable & $UD\overline{U}\overline{D}$ & 483        & -199/\textbf{0} & $\pi^+ + \pi^-$/$\mathbf{\pi}^0 + \mathbf{\pi}^0$ \cr
unstable & $SD\overline{U}\overline{D}$ & 606        & -101/\textbf{0} & $K^0 + \pi^-$/$\mathbf{K}^- + \mathbf{\pi}^0$ \cr
unstable & $SD\overline{S}\overline{D}$ & 728        & 0/\textbf{0} & $K^0 + \overline{K}^0$/$\mathbf{K}^0 + \overline{\mathbf{K}}^0$ \cr
unstable & $SS\overline{U}\overline{U}$ & 733        & -5/\textbf{0} & $K^- + K^-$/$\mathbf{K}^- + \mathbf{K}^-$ \cr
unstable & $SS\overline{U}\overline{D}$ & 731        & -3/\textbf{0} & $K^- + K^0$/$\mathbf{K}^- + \mathbf{K}^0$ \cr
unstable & $SS\overline{S}\overline{S}$ & 2192        & -2/\textbf{0} & $\eta + \eta$/$\mathbf{\eta} + \mathbf{\eta}$ \cr
unstable & $UDD\overline{U}\overline{U}$ & 1223        & 0/\textbf{0} & $p + \pi^- + \pi^-$/$\mathbf{p} + \mathbf{\pi}^- + \mathbf{\pi}^-$ \cr
unstable & $SDU\overline{U}\overline{D}$ & 1499        & -147/\textbf{0} & $\Sigma^+ + \pi^-$/$\mathbf{\Sigma}^0 + \mathbf{\pi}^0$ \cr
unstable & $SSD\overline{U}\overline{D}$ & 1457  & -1/\textbf{0} & $\Xi^0 + \pi^-$/$\mathbf{\Xi}^- + \mathbf{\pi}^0$ \cr
unstable & $SSS\overline{U}\overline{U}$ & 1678  & 0/\textbf{0} & $\Xi^- + K^-$/$\mathbf{\Xi}^- + \mathbf{K}^-$ \cr
 \hline
\end{tabular*}
\label{table:6c}
\end{table}

\begin{table}[h]
\caption{Training set for particles with charmed sakatons}
\begin{tabular*}{\textwidth}{@{\extracolsep{\fill}}lllcc}
 \hline
 Particle & Sakaton & Mass (MeV/$c^2$) & B.E. (MeV) & Products   \cr
         &   structure & calc./\textbf{exp.} & calc./\textbf{exp.} & calc./\textbf{exp.} \cr
 \hline
 $D^0$      & $C\overline{U}$  & 2001/\textbf{1897}  & 1224/\textbf{1328} & $\overline{p} + \Lambda_c^+$/$\overline{\mathbf{p}} + \mathbf{\Lambda}_c^+$ \cr
 $D_s^+$    & $C\overline{S}$  & 2587/\textbf{1968}  & 814/\textbf{1433} & $\overline{\Lambda}^0 + \Lambda_c^+$/$\overline{\mathbf{\Lambda}}^0 + \mathbf{\Lambda}_c^+$ \cr
 $\eta_c$   & $C\overline{C}$  & 3339/\textbf{2980}  & 1231/\textbf{1590} & $\Lambda_c^+ + \overline{\Lambda}_c^+$/$\mathbf{\Lambda}_c^+ + \overline{\mathbf{\Lambda}}_c^+$ \cr
 \hline
  $\Xi_c^0$ & $CS\overline{U}$ & 2606/\textbf{2471}   & 43/\textbf{308} & $\Lambda_c^+ + K^-$/$\mathbf{\Lambda}_c^+ + \mathbf{K}^-$ \cr
   $\Xi_{cc}^+$ & $CC\overline{U}$ & 4092/\textbf{3519}   & 194/\textbf{636} & $\Lambda_c^+ + D^0$/$\mathbf{\Lambda}_c^+ + \mathbf{D}^0$ \cr
   $\Omega_c^0$ & $CSS\overline{U}\overline{D}$ & 2852/\textbf{2698}  & 118/\textbf{268} &
   $\Xi_c^+ + K^-$/$\mathbf{\Xi}_c^+ + \mathbf{K}^-$ \cr
   \hline
unstable & $UD\overline{C}$ & 2944               & -3/\textbf{0} & $p + D^-$/$\mathbf{p} + \mathbf{D}^-$ \cr
unstable & $UU\overline{C}$ & 2945              & -4/\textbf{0} & $p + \overline{D}^0$/$\mathbf{p} + \overline{\mathbf{D}}^0$\cr
unstable & $SU\overline{C}$ & 3121               & -4/\textbf{0} & $\Lambda^0 + \overline{D}^0$/$\mathbf{\Lambda}^0 + \overline{\mathbf{D}}^0$ \cr
unstable & $SS\overline{C}$ & 3705               & -2/\textbf{0} & $\Lambda^0 + D_s^-$/$\mathbf{\Lambda}^0 + \mathbf{D}_s^-$ \cr
unstable   & $CU\overline{D}$ & 2427        & 0/\textbf{0} & $\Lambda_c^+ + \pi^+$/$\mathbf{\Lambda}_c^+ + \mathbf{\pi}^+$ \cr
unstable   & $CU\overline{U}$ & 2523           &  0/\textbf{0} & $\Lambda_c^+ + \pi^0$/$\mathbf{\Lambda}_c^+ + \mathbf{\pi}^0$ \cr
unstable & $CS\overline{S}$   & 3380        &  0/\textbf{0} & $\Lambda_c^+ + \eta$/$\mathbf{\Lambda}_c^+ + \mathbf{\eta}$ \cr
unstable & $CS\overline{C}$   & 4456        & -1/\textbf{0} & $\Lambda^0 + \eta_c$/$\mathbf{\Lambda}^0 + \mathbf{\eta}_c$ \cr
unstable & $CC\overline{S}$ & 4872          & 0/\textbf{0} & $\Lambda_c^+ + D_s^+$/$\mathbf{\Lambda}_c^+ + \mathbf{D}_s^+$ \cr
unstable & $CC\overline{C}$ & 5625          & -1/\textbf{0} & $\Lambda_c^+ + \eta_c$/$\mathbf{\Lambda}_c^+ + \mathbf{\eta}_c$ \cr
unstable & $CU\overline{U}\overline{S}$ & 2365       &  0/\textbf{0} & $D^0 + K^+$/$\mathbf{D}^+ + \mathbf{K}^0$ \cr
unstable & $CU\overline{S}\overline{S}$ & 2953       & -2/\textbf{0} & $D_s^+ + K^+$/$\mathbf{D}_s^+ + \mathbf{K}^+$ \cr
unstable & $CS\overline{S}\overline{S}$ & 3685       & -3/\textbf{0} & $D_s^+ + \eta$/$\mathbf{D}_s^+ + \mathbf{\eta}$ \cr
unstable & $CUD\overline{U}\overline{D}$ & 2568      & -1/\textbf{0} & $\Lambda_c^+ + \pi^+ + \pi^-$/$\mathbf{\Lambda}_c^+ + \mathbf{\pi}^0 + \mathbf{\pi}^0$ \cr
unstable & $CSU\overline{U}\overline{D}$ & 2850      & -102/\textbf{0} & $\Xi_c^0 + \pi^+$/$\mathbf{\Xi}_c^+ + \mathbf{\pi}^0$ \cr
unstable & $CSU\overline{D}\overline{D}$ & 2794 &    -46/\textbf{0} & $\Xi_c^+ + \pi^+$/$\mathbf{\Xi}_c^+ + \mathbf{\pi}^+$ \cr
unstable & $CSS\overline{D}\overline{D}$ & 2970  & 0/\textbf{0} & $\Xi_c^+ + K^0$/$\mathbf{\Xi}_c^+ + \mathbf{K}^0$ \cr
 \hline
\end{tabular*}
\label{table:6d}
\end{table}

\begin{table}[h]
\caption{Low mass states of some mesons with angular momentum quantum numbers $S=L=J=0$. Masses are in MeV/c$^2$.}
\begin{tabular*}{\textwidth}{@{\extracolsep{\fill}}l|ll|ll|ll}
 \hline
Sakaton & $1^1S_0$ & mass & $2^1S_0$ & mass & $3^1S_0$ & mass \cr
structure &          & calc./\textbf{exp.} &  & calc./\textbf{exp.} &  & calc./\textbf{exp.} \cr
 \hline
$D\overline{U}$ & $\pi^-$ & 142/\textbf{140} & $\pi(1300)$ & 1480/\textbf{1300} & $\pi(1800)$ & 1726/\textbf{1816} \cr
$S\overline{U}$ & $K^-$ & 364/\textbf{494} & & 1669 & & 1909 \cr
$S\overline{S}$ & $\eta$ & 1095/\textbf{548} & $\eta(1475)$ & 1974/\textbf{1476} & & 2136 \cr
$C\overline{U}$ & $D^0$ & 2001/\textbf{1897} & & 2942& & 3117 \cr
$C\overline{S}$ & $D_s^+$ & 2587/\textbf{1968} & & 3214 & & 3331 \cr
$C\overline{C}$ & $\eta_c(1S)$ & 3339/\textbf{2980} & $\eta_c(2S)$ & 4282/\textbf{3637} & & 4456 \cr
\hline
\end{tabular*}
\label{table:4.1e}
\end{table}

The resulting masses of  hadrons are shown in the third column of tables \ref{table:6c} and \ref{table:6d}.  The binding energies (B.E.) are in the 4th column and the lowest-energy dissociation products are in the 5th column.

Ideally, the binding energies of unstable tetrasakatons and baryons\footnote{They are shown in the lower portions of Tables \ref{table:6c} and \ref{table:6d}.}  must be equal to zero. In practice this can be achieved only with very large and diffuse basis sets, which allow the wave functions of dissociation products to separate widely, so that their repulsion is minimized. For computational reasons our basis sets were limited. This explains why  some residual repulsion (reflected in negative binding energies from 0 to -5 MeV) remained for several dissociated unstable species. Extremely large negative binding energies of $UD\overline{U}\overline{D}$,  $SD\overline{U}\overline{D}$, $SDU\overline{U}\overline{D}$, $CSU\overline{U}\overline{D}$ and $CSU\overline{U}\overline{D}$ are explained by the fact that they have converged to metastable  dissociated configurations (=local minima) $U\overline{U}+D\overline{D}$,  $S\overline{U}+ D\overline{D}$, $SD\overline{D} +U\overline{U}$, $CS\overline{D} +U\overline{U}$, and  $C+S\overline{U} + U\overline{D}$, respectively.

The next step is to consider properties that have not been used directly in the fitting. First, we looked at the two charmed baryons whose existence is predicted by the quark model and whose experimental confirmation is still lacking. These are the $\Omega_{cc}^+ $ and  $\Omega_{ccc}^{++} $ particles. We found that $\Omega_{cc}^+ (= CCS\overline{U}\overline{D})$ is stable with the mass of 4430 MeV/$c^2$ and binding energy of 25 MeV with respect to the $\Omega_{cc}^{+} \to \Xi_{cc}^+ + K^0 (= CC\overline{U} + S\overline{D})$ dissociation channel.  The calculated mass of $\Omega_{ccc}^{++} (= CCC\overline{U}\overline{D})$ is 6099 MeV/$c^2$, which means that this particle dissociates spontaneously as $\Omega_{ccc}^{++} \to \Xi_{cc}^+ + D^+ (= CC\overline{U} + C\overline{D})$.

Next we verified that all 133 possible tetrasakaton ($\sigma \sigma \overline{\sigma}\overline{\sigma}$) and baryon ($\sigma \sigma \overline{\sigma}$ and $\sigma \sigma \sigma \overline{\sigma}\overline{\sigma}$) species not presented in Tables \ref{table:6c} and \ref{table:6d} are unstable in our approach, as expected.

Other interesting pieces of information, which have not been involved in the fitting, are the meson excitation energies. Note that the strongly attractive  $\sigma \overline{\sigma}$ potential (see Fig. \ref{fig:3}) can accommodate a few stationary states that can be regarded as excitations of the ground-state meson.
In Table \ref{table:4.1e} we show calculated masses of 3 lowest spherically symmetric  ($J^P=0^-$) meson states and compare them with experimental numbers where available. The same basis set was used for the ground and excited states. Obtained excitation energies of the order of several hundreds of MeV are roughly consistent with observed data. This gives us some confidence regarding the overall shape of the selected interaction potentials.

\section{Discussion}
\label{sc:discussion}

The most important lesson of the above calculations is that the Sakata model is \emph{qualitatively} correct, at least in the part concerning masses of strongly interacting particles. With properly adjusted interaction potentials, this model correctly predicts the stability of those species, which are found stable in nature. On the other hand, the unbound combinations of sakatons are exactly those, which were not seen in experiments. The calculated masses of stable particles (see Tables \ref{table:6c} and \ref{table:6d}) sometimes differ from experimental values by hundreds of MeV/c$^2$. For example, masses of baryons are systematically overestimated. However, such discrepancies are expected due to our use of simplified 2-particle potentials (\ref{eq:potential}). One can expect that true sakaton interactions have a more sophisticated form.

For example, in our approach, $U-D$,  $U-U$ and $D-D$  potentials are purely repulsive. This does not allow us to describe bound states like $UD$ (deuteron) or $UUDD$ ($\alpha$-particle). It seems plausible that these interactions (especially the $U-D$ potential) can be slightly modified so as to make them attractive at distances $\approx$1-2 fm (see broken line in Fig. \ref{fig:3}). Then it might be possible to reproduce the bonding of protons and neutrons in nuclei. Such a possibility is especially exciting as it would allow us to describe the stabilities of mesons, baryons and nuclei within the same set of sakaton interactions.

Another missing piece is the absence of relativistic corrections that may
include momentum-dependent, spin-orbit, spin-spin, and contact
interactions. It is well-established that they can contribute up to several hundreds of MeV to the overall energy balance of mesons and nuclei.

One can also add to (\ref{eq:potential}) terms which change the number and/or types of particles. For example, terms like\footnote{Here $u,\overline{u}, d, \overline{d}$ are annihilation operators for the $U, \overline{U},D, \overline{D}$ sakatons, and $u^{\dag},\overline{u}^{\dag}, d^{\dag}, \overline{d}^{\dag}$ are their creation operators.}

\begin{eqnarray}
V_{mix} \propto u^{\dag}\overline{u}^{\dag}d\overline{d} + d^{\dag}\overline{d}^{\dag}u\overline{u} + \ldots
\label{eq:V4}
\end{eqnarray}

\noindent are responsible for the mixing of $U\overline{U}$ and $D \overline{D}$ states and for the mass splitting between $\pi^0 = 1/\sqrt{2}(U\overline{U} - D \overline{D})$ and $\eta' \approx 1/\sqrt{2}(U\overline{U} + D \overline{D})$ mesons \cite{PDBook}. Without interaction (\ref{eq:V4}) particles $\pi^0$ and $\eta'$ have the same mass, while experimentally their masses are quite different: $m(\pi^0)=$\textbf{135} MeV/$c^2$, $m(\eta')=$\textbf{958} MeV/$c^2$. This indicates the significant role of terms like (\ref{eq:V4}). Generally, one can also expect the presence of interactions that lead to the mixings $N\overline{N} \leftrightarrow S\overline{S} \leftrightarrow C\overline{C} $. Our neglect of these interactions may partially explain the overestimation of masses of $\pi^0$, $\eta$, and $\eta_c$ mesons.

One may argue that Sakata's assumption of a fundamental point-like proton must be wrong because, being probed by truly point-like electrons, the proton demonstrates a sizeable charge radius of 0.877 fm. However, this experimental fact can be accommodated within the Sakata model as well. To achieve that, one can assume the presence of particle-number-changing  interaction terms like

\begin{eqnarray}
V_{unphys} \propto \overline{u}^{\dag}u^{\dag}u^{\dag}u + \overline{d}^{\dag}d^{\dag} u^{\dag}u + u^{\dag}\overline{u}uu + u^{\dag}\overline{d}du +\ldots
\label{eq:V5}
\end{eqnarray}

\noindent in the Hamiltonian. In the classification of \cite{mybook} these terms are called ``unphys''. If they are present, then single ``bare'' proton states $u^{\dag}| 0 \rangle$ are not eigenstates of the total Hamiltonian. To make the theory sensible, one needs to perform a renormalization. If coefficient functions in the interaction (\ref{eq:V5}) are properly chosen,\footnote{e.g., if they decay rapidly at large values of momenta; see Theorem 7.12 in \cite{mybook}} then all loop integrals are finite, the renormalization effects are finite too, and the ``bare'' proton becomes ``dressed'' by a cloud of virtual pairs and pions, thus acquiring a non-zero size \cite{Grobe}.

In spite of the deficiencies listed above, our results indicate a remarkable consistency between the quark and Sakata models: both models predict the same set of stable hadron states.\footnote{This does not apply to the $\Omega_{ccc}^{++}$ particle whose  quark content is $ccc$. This particle appears unstable in our approach. The experimental confirmation of its existence is still lacking.}  This suggests that Sakata's idea about the hadron structure has a non-vanishing fighting chance against the quark model. Further studies with more elaborate potentials would be certainly welcome. In addition to the masses of stable species considered here, these future studies should address resonances and scattering properties as well.

I would like to thank Dr. Robert Wagner for reading the manuscript and for helpful critical comments.

%\newpage

%\bibliographystyle{plunsrt}
%\bibliography{../xbib}

\end{document}